\documentclass[aps,prl,a4paper,superscriptaddress,floatfix,showpacs,notitlepage,preprint]{revtex4-2} 

\usepackage{soul,xcolor}
\usepackage{amsmath}
\usepackage{amsfonts}
\usepackage{amssymb}
\usepackage{physics}
\usepackage{wrapfig, blindtext}
\usepackage{graphicx}
\usepackage{cancel}
\usepackage{verbatim}
\graphicspath{{./figs/}}

\usepackage{cleveref}
\crefname{equation}{}{}
\Crefname{equation}{}{}

\newcommand{\nb}[1]%
{\begingroup{\color{red} (fpm: #1)}\endgroup}

\newcommand{\nbold}[1]%
{\begingroup{\color{blue} (fpm (DONE): #1)}\endgroup}

\newcommand{\nbjm}[1]%
{\begingroup{\color{red} (JM: #1)}\endgroup}

\newcommand{\nboldjm}[1]%
{\begingroup{\color{blue} (JM (DONE): #1)}\endgroup}

\begin{document}

\title{De Haas--van Alphen effect in graphene}

\author{Juuso Manninen}
\thanks{These two authors contributed equally}
\affiliation{Low Temperature Laboratory, Department of Applied Physics, Aalto University, PO Box 15100, FI-00076 AALTO, Finland}
 \affiliation{
  QTF Centre of Excellence, Department of Applied Physics, Aalto University, PO Box 15100, FI-00076 AALTO, Finland}
\author{Antti Laitinen}
\thanks{These two authors contributed equally}
\affiliation{Low Temperature Laboratory, Department of Applied Physics, Aalto University, PO Box 15100, FI-00076 AALTO, Finland}
\affiliation{Department of Physics, Harvard University, Cambridge, MA 02138, USA}
\author{Francesco Massel}
\affiliation{Department of Physics, Nanoscience Center, University of Jyv\"{a}skyl\"{a}, FIN 40014, Finland}
\affiliation{Department of Science and Industry Systems, University of
  South-Eastern Norway, PO Box 235, Kongsberg, Norway}
\author{Pertti Hakonen}
\email[]{pertti.hakonen@aalto.fi}
\affiliation{Low Temperature Laboratory, Department of Applied Physics, Aalto University, PO Box 15100, FI-00076 AALTO, Finland}
 \affiliation{
  QTF Centre of Excellence, Department of Applied Physics, Aalto University, PO Box 15100, FI-00076 AALTO, Finland}

\begin{abstract}
  In our work, we study the dynamics of a graphene Corbino disk
  supported by a gold mechanical resonator in the presence of a
  magnetic field. We demonstrate here that our graphene/gold
  mechanical structure exhibits a nontrivial resonance frequency
  dependence on the applied magnetic field, showing how this feature
  is indicative of the de Haas--van Alphen effect in the graphene
  Corbino disk.  Our findings are the first evidence of dHvA
  effect for massless Dirac fermions. By relying on the mechanical resonances of the Au structure, our detection scheme is
  essentially independent of the material considered and  
  can be applied for dHvA measurements on any
  conducting 2D material. In particular, the scheme is expected to be an important tool in studies of centrosymmetric transition metal dichalcogenides (TMDs) crystals, shedding new light on hidden magnetization and interaction effects.
\end{abstract}

\maketitle

\section{Introduction}
\label{sec:intro} 

As theoretically shown by Landau and Peierls in the 1930s
\cite{Landau1930,Peierls1933}, the de Haas--van Alphen (dHvA) effect
consists of a periodic oscillation of the magnetization (and the
magnetic susceptibility) as a function of the magnetic field. Along with other 
magnetic-field-induced phenomena, such as the Shubnikov de Haas (SdH) conductance oscillations, 
the quantum Hall effect, and quantum capacitance oscillations, the origin of the dHvA effect is 
traced back to the modification of the electronic spectrum in the presence of a magnetic field. 
Since in this case electronic motion becomes quantized due to the formation of Landau 
levels, which are ultimately responsible for the non-trivial properties of the considered 
electronic system, it is quite natural that
the dHvA
effect has served as the central probe in studies of the shape of the Fermi
surface in normal metals.

Besides investigations of the dHvA effect in conventional materials, magnetic properties of 
two-dimensional (2D) materials have been investigated actively \cite{Ezawa2013,Dolgopolov2014}. 
Unlike the 3D case, where the field
dependence of the magnetization is described by the classical 3D
Landau-Kosevich formula, for 2D samples the magnetization shows a
characteristic sawtooth pattern both for massive and massless Dirac
fermions~\cite{Lukyanchuk.2011}. On the experimental side, in 2D the dHvA effect was first observed by
Eisenstein et al. in 1985~\cite{Eisenstein1985} in a 2D electron gas
(2DEG), while a clear sawtooth pattern for the magnetization vs.\
inverse magnetic field predicted in \cite{Peierls1933} was resolved about ten years
later~\cite{Wiegers1997}. Focusing on detection techniques of magnetic properties based on mechanical
motion, surface acoustic waves (SAW) have extensively been used for
imaging of integer and fractional quantum Hall
states (QH)~\cite{Wixforth1989} in conventional GaAs 2DEG systems. In
SAW-based techniques, mechanical motion is coupled to the electron
system due to piezoelectric response of GaAs, and variation in the
compressibility of the electron system
modulates attenuation and sound velocity in the material. Apart from
SAW resonances, QH states in a 2DEG have also been
investigated through curling \cite{Okamoto2014}, cantilever
\cite{Harris1999, Bleszynski-Jayich2009}, and torsional modes
\cite{Eisenstein1985}.

For two-dimensional materials, such as graphene and transition metal dichalcogenides (TMDs), while
SdH conductance oscillations have been 
reported \cite{Novoselov2005,Zhang2005,Tan2011} in graphene, experiments have 
not yet revealed the dHvA effect. 
This is the focus of our work: we report here, for the
first time, a dHvA measurement for a graphene membrane in a Corbino
geometry coupled to a gold (Au) mechanical resonator. This
configuration allows us to exploit the possibilities offered by suspended 
resonators in addressing the magnetic properties of Dirac
fermions in graphene and, in principle, the carrier-dependent magnetic behavior in other 2D materials. 

While previous experiments on suspended graphene mechanical
resonators have focused on displaying the periodic variation of
quantum capacitance as a function of magnetic field $B$ in the integer
QH regime~\cite{Singh2012,Chen2016}, in our work, we extend such
principle, showing how it is possible to perform a mechanical readout
not only of the quantum capacitance, but also of the magnetic
susceptibility of the graphene sheet. This can be done evaluating the
explicit functional dependence of the quantum capacitance on the
magnetic susceptibility.

\begin{figure}[tbp]
  \includegraphics[width=0.42\textwidth]{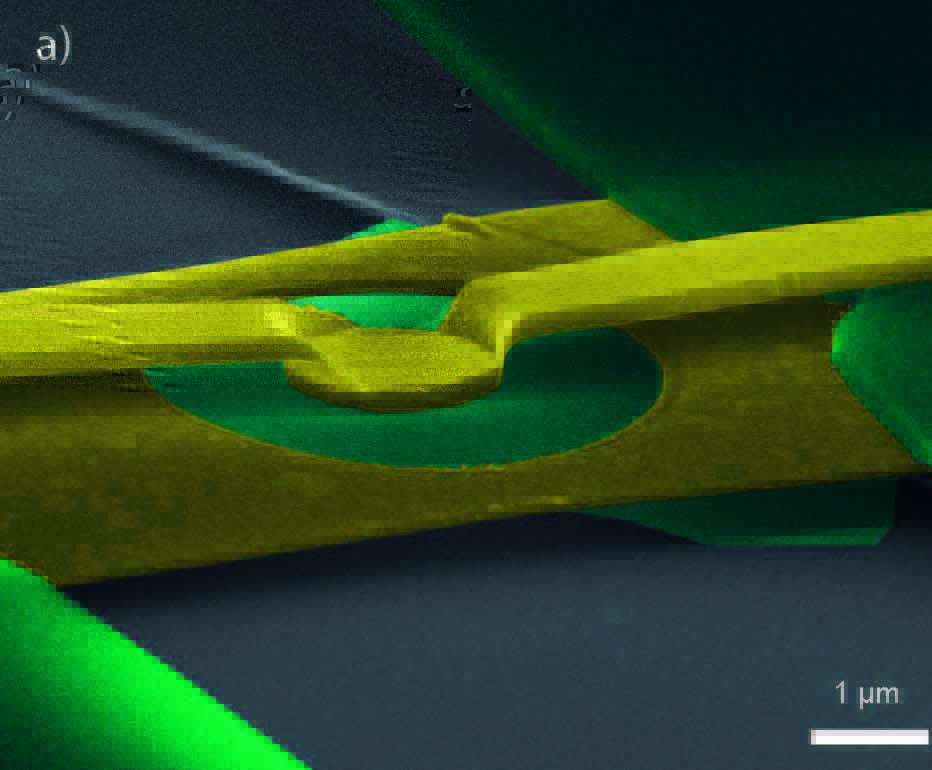}
  \includegraphics[width=0.57\textwidth]{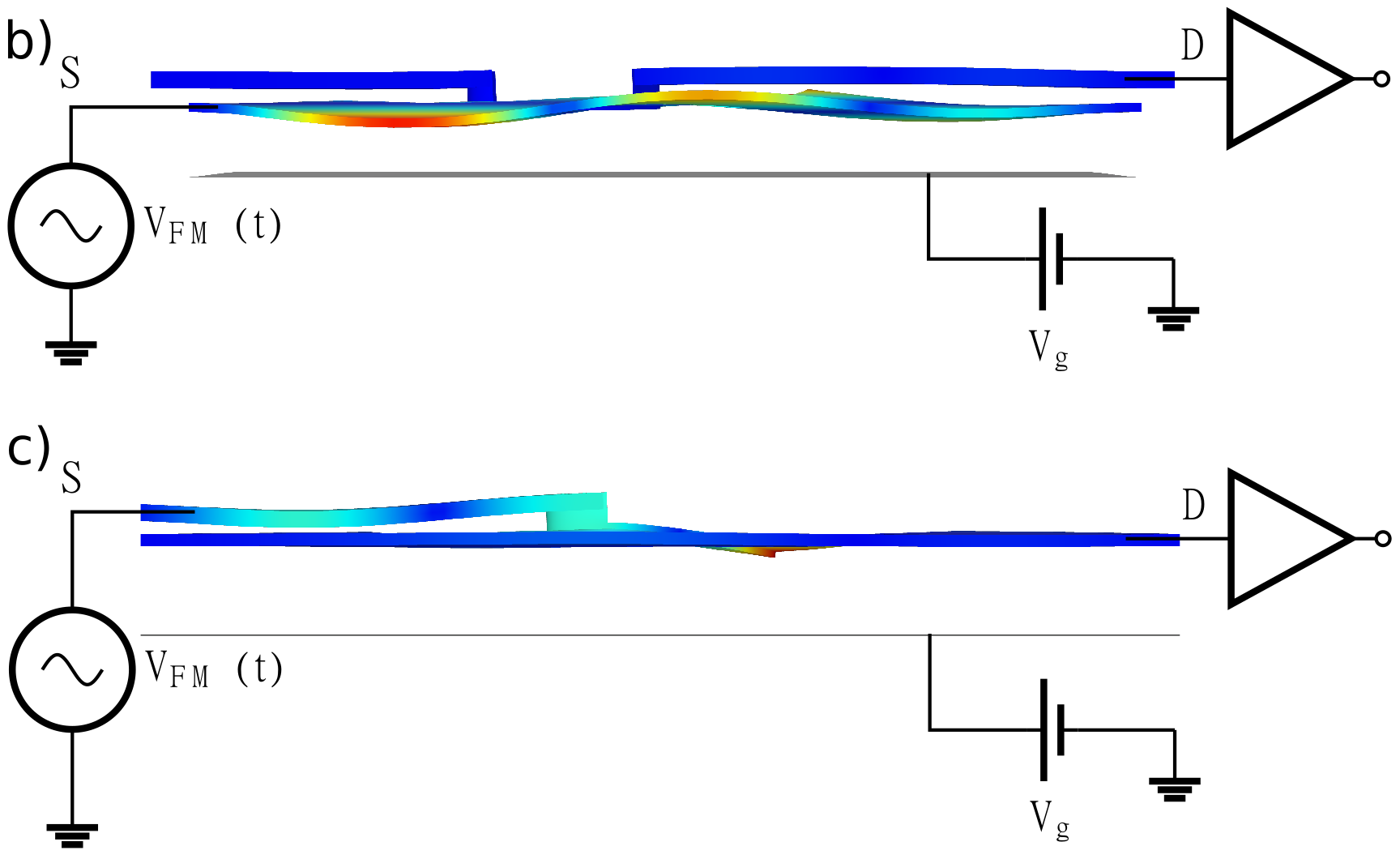}
  \caption{ {\bf Sample structure and key mechanical modes.} a) SEM image of the measured device B2. The ring shaped
    graphene colored green, Au parts appear as yellow, polymer support
    as dark green, and the substrate is gray. The length of the lower gold beam amounts 8 $\mu$m. 
    Schematic of our measurement method for the device b) B2 and c) B1.5 
    consisting of two Au electrodes, one graphene Corbino disk, a back gate 
    voltage $V_\mathrm{g}$, and the frequency modulation voltage 
    $V_\mathrm{FM}$. Mode shapes (not in scale) obtained from FEM 
    simulations of the respective 
    devices utilized in this study are depicted with a color gradient 
    highlighting the physical displacement.}
   \label{fig:schematic}
\end{figure}

The basic measurement principle of our experiment is illustrated
schematically in Fig.~\ref{fig:schematic}. Our experiments were
carried out on two devices: B2 and B1.5. Device B2 consists of \emph{two Au beams}, one
graphene Corbino disk, and a back gate to which a voltage
$V_\mathrm{g}$ is applied, controlling the charge density $n$ on the
graphene disk. The Corbino disk couples the
two Au beams together mechanically; the parallel Au beams are located
at different heights, about 150~nm apart, supported by a bend in the
center of the upper Au beam  (Fig.~\ref{fig:schematic}a,c). Device B1.5 consists of 
\emph{one and a half Au beams}:  the top Au beam has been replaced by a gold cantilever in it (Fig.~\ref{fig:schematic}b).  An electrical current
can be passed through between the terminal source and drain, marked in
Fig.~1b,c by S and D, respectively.
The Au beams and the cantilever have nearly the same bending rigidity. 
This leads to the appearance of mechanical resonance frequencies
between 10 and 100 MHz for both devices, basically governed by the gold structures, with 
the dynamics of the graphene membrane being dictated by
the motion of the graphene/gold boundary conditions.

Owing to the mechanical motion of
the graphene membrane, the charge distribution within the Corbino disk
will be affected, altering the capacitive forces at play and,
consequently, the mechanical resonances of the combined structure in the
presence of a external biasing voltages and/or magnetic fields.
In our devices, we can picture the gold
oscillator acting as a mechanical detector of the magnetic
(magnetization, magnetic susceptibility) and electronic (quantum
capacitance) properties of the graphene disk.
Our approach is along the lines of cantilever
sensing~\cite{Bleszynski-Jayich2009}, but differs fully in the sense
that our method allows us to investigate different materials independent
of the probing mechanical cantilever (in our case the Au structure) 
and thus the method is applicable to any 2D material, many of which can 
be fabricated into mechanical resonators \cite{Lemme.2020,Siskins.2020}. 
The idea of using mechanical motion to measure the magnetic properties
of a graphene membrane introduced here can be considered as a part of the 
emerging field of sensing with 2D mechanical resonators \cite{Steeneken.2021}.

\section{Basic experimental characteristics}

Owing to the difference in effective mass between the Au and graphene
portions of the devices, two basic types of resonances were observed
in our samples: (low-frequency 10-40 MHz) combined gold-graphene modes (``Au mode'') and
(high-frequency $\gtrsim$ 90 MHz) pure graphene resonances in the
Corbino disk \cite{Kamada.2021}.
Due to the mechanical properties of the Au beams, for 
combined gold-graphene modes, there is a wider range of 
driving fields for which the linear detection of the 
quantum Hall states in graphene is possible in comparison 
with pure graphene modes. For these modes, the linear regime 
is limited to oscillation amplitudes around $100$ pm \cite{Song2012}. 
For this reason, and for the ``material independence'' of Au modes, 
in the following we focus on the latter ones.

\begin{figure}[tbp]
\includegraphics[width=1\textwidth]{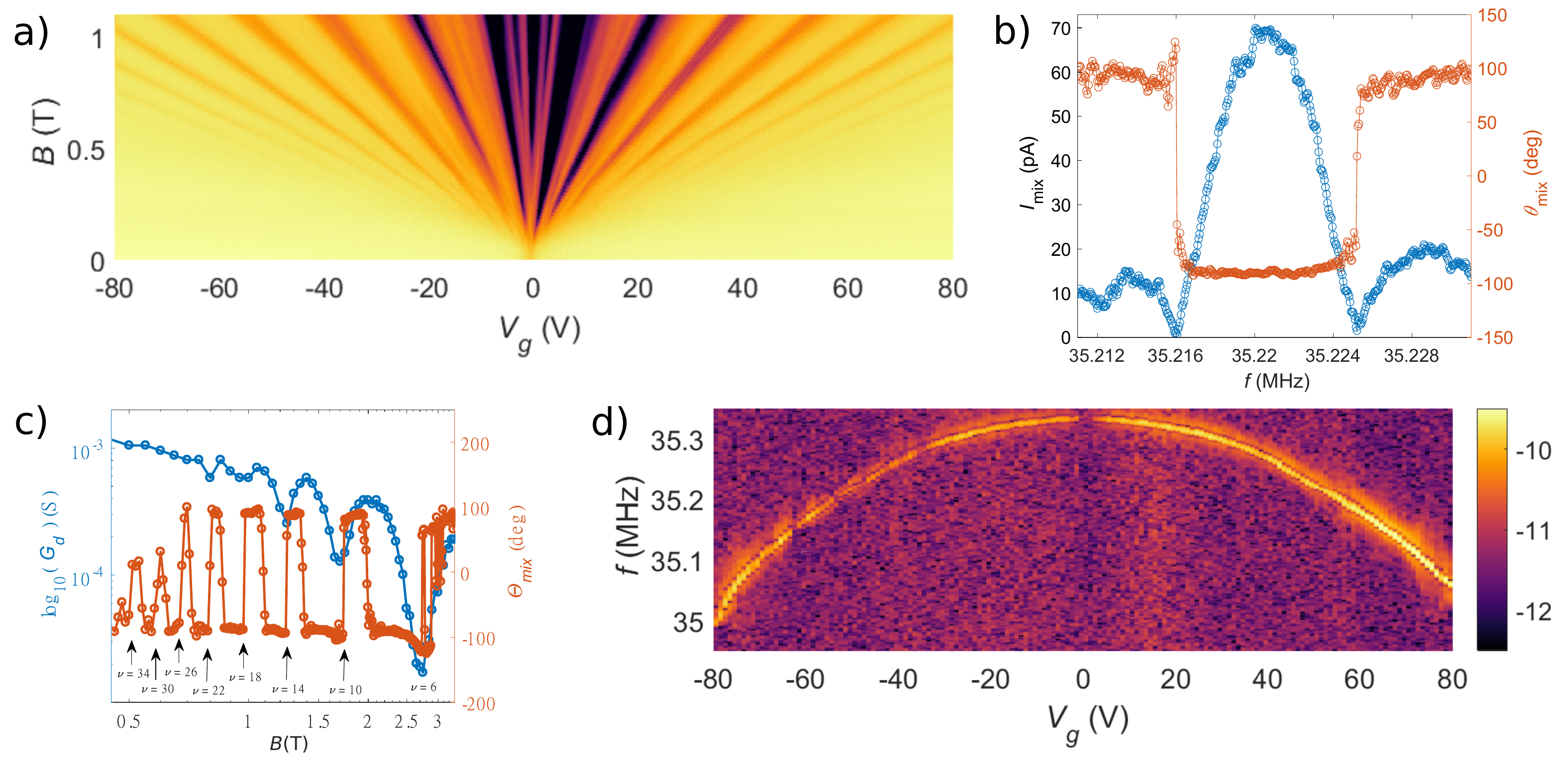}
\caption{{\bf Characteristics of the suspended graphene Corbino disk B2.} 
  a) Landau fan diagram over the gate
  voltage $V_\mathrm{g}$ vs. the magnetic field $B$ plane, measured at $V_\mathrm{g} = 60$ V ($n=4.3\cdot 10^{11}$ cm$^{-2}$). 
  b) Mixing current $I_\mathrm{mix}$ and its phase of the 35 MHz
  resonance in zero magnetic field. 
  c) Logarithm of
  conductance and the mixing current phase at resonance maximum as a function of magnetic
  field.
  d) $V_\mathrm{g}$
  dependence of the logarithm of mixing current $\log_{10}(I_\mathrm{mix})$
  of the 35 MHz resonance. }
\label{fig:sample}
\end{figure}

The quality of the investigated graphene disks with appreciable built-in strain was assessed by
measuring the Landau fan diagram, $G(V_\mathrm{g},B)$, which is illustrated in
Fig. \ref{fig:sample}a for sample B2 at $n=4.3\cdot 10^{11}$ cm$^{-2}$.
The dark regions denote large resistance that correspond to hopping
conductance along localized levels within gapped QH
states. A full set of integer QH states becomes visible at
fields $B \geq 0.5$ T, while the fractional QH state
$\nu=1/3$ appears starting from $B\approx 3$ T, similar to works in
Refs.~\citep{Laitinen2018_halffilling,Kumar2018}.

The mechanical resonance properties of the samples were investigated using the FM mixing
technique \cite{Gouttenoire2010,Eichler2011}, which was employed due
to the clear-cut form of the mixing signal, exhibiting a sharp and consistent three-lobed 
peak structure, see Fig.~\ref{fig:sample}b.
Furthermore, both the conductance and the phase of the mixing current $I_\mathrm{mix}$ at mechanical the resonance were found to reflect the non-trivial $B$ dependence of the electronic properties (see
Fig. \ref{fig:sample}c): the local minima in $G$ coincide with the
upwards phase flips in $I_\mathrm{mix}$ caused by the change of sign of the derivative
$dG/dV_\mathrm{g}$ across the gapped QH states. Thus, the phase flips in $I_\mathrm{mix}$
can be employed as sensitive detectors of the QH states in
suspended graphene, and a Landau level sequence up to $\nu=34$ can be resolved in Fig. \ref{fig:sample}c.

\section{Forces acting on the  resonator}

Our graphene/gold mechanical resonator can be described as a capacitor
with one movable plate coupled to an external voltage source. In
addition to the conventional electromagnetic field energy between the
capacitor plates, the system exhibits a contribution to its total
energy deriving from the finite density of states (DOS) of graphene.
In other terms, the external voltage $V_\mathrm{g}$ provides the
external work required to build up the field between the capacitor
plates, as in conventional capacitors, and, concomitantly, it is
needed to overcome the extra energy associated with the finite DOS of
graphene.

Such energy contribution leads to a reduction of the force
between the plates of the movable capacitor
\begin{align}
   F = \frac{1}{2} C_\mathrm{g}' \left( V_\mathrm{g} - \frac{\mu}{e} \right)^2,
  \label{eq:1}
\end{align}
where $V_\mathrm{g}$ is the external applied voltage, $\mu$ the
graphene chemical potential, $C_\mathrm{g}$ the (position-dependent)
geometric capacitance of the structure, and $C'_\mathrm{g}$ its
derivative with respect to the displacement of the graphene/gold
electrode (see Supplementary Information). 
This formula fits well the measured gate
dependence of the lower Au beam resonance presented in
Fig. \ref{fig:sample}d.

In the presence of an external magnetic field, the resonant frequencies of our structures $f_\mathrm{n}$ exhibit a nontrivial dependence on $B$ reflecting the emergence of Landau levels in the spectrum of graphene. Along the lines of Ref.~\cite{Chen2016}, we first express this frequency change
in terms of quantum capacitive effects
\begin{align}
  \Delta f_{\mathrm{n}}\doteq f_{\mathrm{B,n}}-f_{\mathrm{B=0,n}}=
  -\frac{\partial f_\mathrm{0,n}}{\partial V}\left[\frac{\Delta \mu}{e}+
  C_\mathrm{g} V_\mathrm{g} \Delta\frac{1}{C_\mathrm{q}} \right]
  +\frac{f_\mathrm{0,n}}{2 k_\mathrm{0,n}}
  \left(C'_\mathrm{g} V_\mathrm{g}\right)^2\Delta\left(\frac{1}{C_\mathrm{q}}\right) ,
  \label{eq:2}
\end{align}
where $\partial f_\mathrm{0,n}/\partial V$ is the $B=0$ resonant
frequency tunability with the gate voltage due to tensioning effects,
$C_\mathrm{q}= e^2 \partial n/\partial \mu$ is the graphene quantum
capacitance, and $k_\mathrm{0,n}=4 \pi^2 f_\mathrm{0,n}^2 \rho$ is the
effective elastic constant of mode $\mathrm{n}$ at $B=0$ (see
Supplementary Information).  We note here that, due to the large
effective elastic constant related to the density of gold taking part in the 
mechanical motion, in our setup the last term of Eq.~\eqref{eq:2} is
negligible, leading to the appearance of local minima in the magnetic
field dependence of $f_\mathrm{n}$.

\section*{Quantum capacitance and de Haas--van Alphen effect in graphene}

Central to our analysis is the notion that it is possible to establish a direct
relation between the graphene quantum capacitance $C_\mathrm{q}$ \cite{Ponomarenko2010,Yu2013} and the
magnetic susceptibility $\chi_\mathrm{m}$. This correspondence allows
us to relate the frequency shifts described in Eq.~\eqref{eq:2}
to the magnetic susceptibility shifts and therefore consider
$ \Delta f_{\mathrm{n}}$ as a reliable measure of the dHvA effect.

As mentioned in relation to the force acting on the graphene/gold
resonator given in Eq.~\eqref{eq:1}, the properties of the whole
system (moveable capacitor + graphene disk) can be derived from the
relevant thermodynamic potential.  In our case, given the
$\bar{\mu}\doteq e V_\mathrm{g}=$const constraint, we consider the
grand canonical potential $\Omega(\bar{\mu},B)$, where the
electrochemical potential $\bar{\mu}=e V_\mathrm{g}$ is the global
control parameter. From this perspective, $\mu$ is an energy
contribution associated with the finite DOS of graphene and thus does
not represent an independent control parameter and will therefore, in
general, depend on the external magnetic field.

If we now confine ourselves to the analysis of the graphene sheet,
i.e. we exclude the field between resonator and backgate from the
definition of the system, we can assume that the independent control
parameter is $\mu$. This allows us to write the thermodynamic
potential associated with the Corbino disk as
$\Omega_\mathrm{disk}(\mu,B)=\Omega_\mathrm{0}(\mu)+\Omega_\mathrm{osc}(\mu,B)$.
The oscillatory dependence on $B$ is a direct consequence of the
appearance of Landau levels in the energy
spectrum~\cite{Peierls1933,Sharapov2004,Katsnelson2012}.  From
$\Omega_\mathrm{osc}$, it is possible to calculate the oscillating
part of the magnetic susceptibility $\chi_\mathrm{m,osc}$
\begin{align}
  \label{eq:3}
 \chi_\mathrm{m,osc} &= -\frac{N e  \mu} {4 \pi \hbar B}
    \frac{}{}
  \frac{\left( \mu^2 - \frac{\gamma^2}{4} \right)
   \left\{
   e^{- 2 \pi \gamma \mu /\hbar \omega_\mathrm{D}}-\cos\left[\frac{2 \pi}{(\hbar \omega_\mathrm{D})^2}
   \left( \mu^2 - \frac{\gamma^2}{4} \right)\right]
  \right\}
  +\gamma \mu
  \sin\left[
  \frac{2 \pi}{(\hbar \omega_\mathrm{D})^2}\left( \mu^2 - \frac{\gamma^2}{4} \right)
  \right]
  }
  {
  (\hbar \omega_\mathrm{D})^2 \left\{\cosh \left[ \frac{2 \pi \gamma \mu }
 {\left(\hbar \omega_\mathrm{D}\right)^2}\right]
 -\cos\left[\frac{2 \pi}{(\hbar \omega_\mathrm{D})^2}\left( \mu^2 - \frac{\gamma^2}{4} \right)\right]\right\}
  }
\end{align}
with $N$ as the spin degeneracy factor, $\omega_\mathrm{D}=\sqrt{2 e B /\hbar}$ and
$\gamma = \hbar/(2 \tau_q)$,
where $\tau_q$ is the quantum
scattering time (see Supplementary Information and
Ref.~\cite{Sharapov2004}).
  
The connection between the description given by $\Omega(\bar{\mu},B)$
and $\Omega_\mathrm{disk}(\mu,B)$ can be understood as though
the external control parameter $\bar{\mu}$ determines, along with $B$,
the control parameter of the graphene disk. More specifically, the
magnetic field dependence of $\mu$ is a consequence of the finite
value of the geometric capacitance $C_\mathrm{g}$, since
$\bar{\mu}=e^2 n(B) / C_\mathrm{g}+\mu$. If we ideally think of
removing the moveable capacitor by directly applying the voltage 
to the graphene sheet, i.e. $C_\mathrm{g} \rightarrow \infty$, 
the two control parameters
coincide, implying that the dependence of $\mu$ on $B$,
for finite $C_\mathrm{g}$, can be interpreted 
as a feature of the measurement setup and not an intrinsic
property of the material sample (graphene in our case).  Similarly,
the measurement of thermodynamic quantities associated with
$\Omega_\mathrm{osc}(\mu,B)$ (such as
$\chi_\mathrm{m,osc}= e^2 \partial M_\mathrm{osc}/\partial B$ and
$C_\mathrm{q,\,osc}=e^2 \partial n_\mathrm{osc}/\partial \mu$)
represent a direct probe of the properties of the sample.

The general observation allowing to relate the frequency shift given
in Eq.~\eqref{eq:2} to the dHvA effect, is that both the quantum
capacitance
$C_\mathrm{q,osc}=e^2 \partial n_\mathrm{osc}/\partial \mu$
and the magnetic susceptibility oscillations
$\chi_\mathrm{m,osc}=e^2 \partial M_\mathrm{osc}/\partial B$
are a manifestation of the same underlying phenomenon, i.e. the Landau
level structure of graphene, ultimately described by the
thermodynamic potential $\Omega_{\mathrm{disk}}(\mu,B)$. For the case of
graphene,  we can establish
a  relation between $C_\mathrm{q,osc}$ and $\chi_\mathrm{m,osc}$,
\begin{align}
   C_\mathrm{q,osc} = \frac{\chi_\mathrm{m,osc}}{\Gamma(\mu,B)},
  \label{eq:5}
\end{align}
where, for $\mu \gg \gamma$, $\Gamma(\mu,B)=\left(\frac{\mu}{ 2 e B}\right)^2$
(see Supplementary Information). Equation~\eqref{eq:5} directly
translates the quantum capacitance oscillations with the magnetic
field into oscillations of the magnetic susceptibility. While the
analytical relation~\eqref{eq:5} is specific to graphene, it is
possible to apply the same idea to other materials: the analytical
structure of $\Omega_\mathrm{disk}(\mu,B)$ implies that the origin of
the oscillatory behavior of $C_\mathrm{q}$ and $\chi_\mathrm{m}$ is
shared and it is therefore possible to relate the two.

From Eq.~\eqref{eq:5}, considering the case of negligible $\Delta \mu$ 
it is also possible
to express the frequency shift given in Eq.~\eqref{eq:2} as a function
of $\chi_\mathrm{m}$ as
\begin{align}
  \Delta f_{\mathrm{n}}=\left[
    \frac{\partial f_{0,\mathrm{n}}}{\partial V} C_\mathrm{g} V_\mathrm{g}
    -\frac{f_\mathrm{0,n}}{2 k_\mathrm{0,n}}\left(C'_\mathrm{g} V_\mathrm{g}\right)^2 
  \right]
    \frac{
     \chi_\mathrm{m,osc}/\Gamma(\mu,B)}
     {C_\mathrm{q,0}
     \left[
     C_\mathrm{q,0} +  
    \chi_\mathrm{m,osc}/ \Gamma(\mu,B)
     \right]}, 
  \label{eq:6}
\end{align}
where
$C_\mathrm{q,0}=e^2 \partial n_0/ \partial \mu=N e^2\mu/(\pi v_\mathrm{F}^2 \hbar^2)$,
with the spin degeneracy factor $N=2$. The relation given in
Eq.~\eqref{eq:6} allows us to infer the oscillations of the magnetic
susceptibility, characteristic of the dHvA effect, directly from the
frequency shift measurement in the graphene Corbino disk.
Furthermore, Eq.~\eqref{eq:6}
allows us to establish the optimal value of the gate voltage
$V_\mathrm{g}= \frac{C_\mathrm{g} k_\mathrm{0,n}}{C'_\mathrm{g} f_\mathrm{0,n}} \frac{\partial f_{0,\mathrm{n}}}{\partial V}$
leading to the maximum frequency shift.

\section{Quantum Hall states and mechanical resonances}

\begin{figure}[h]
\includegraphics[width=1\textwidth]{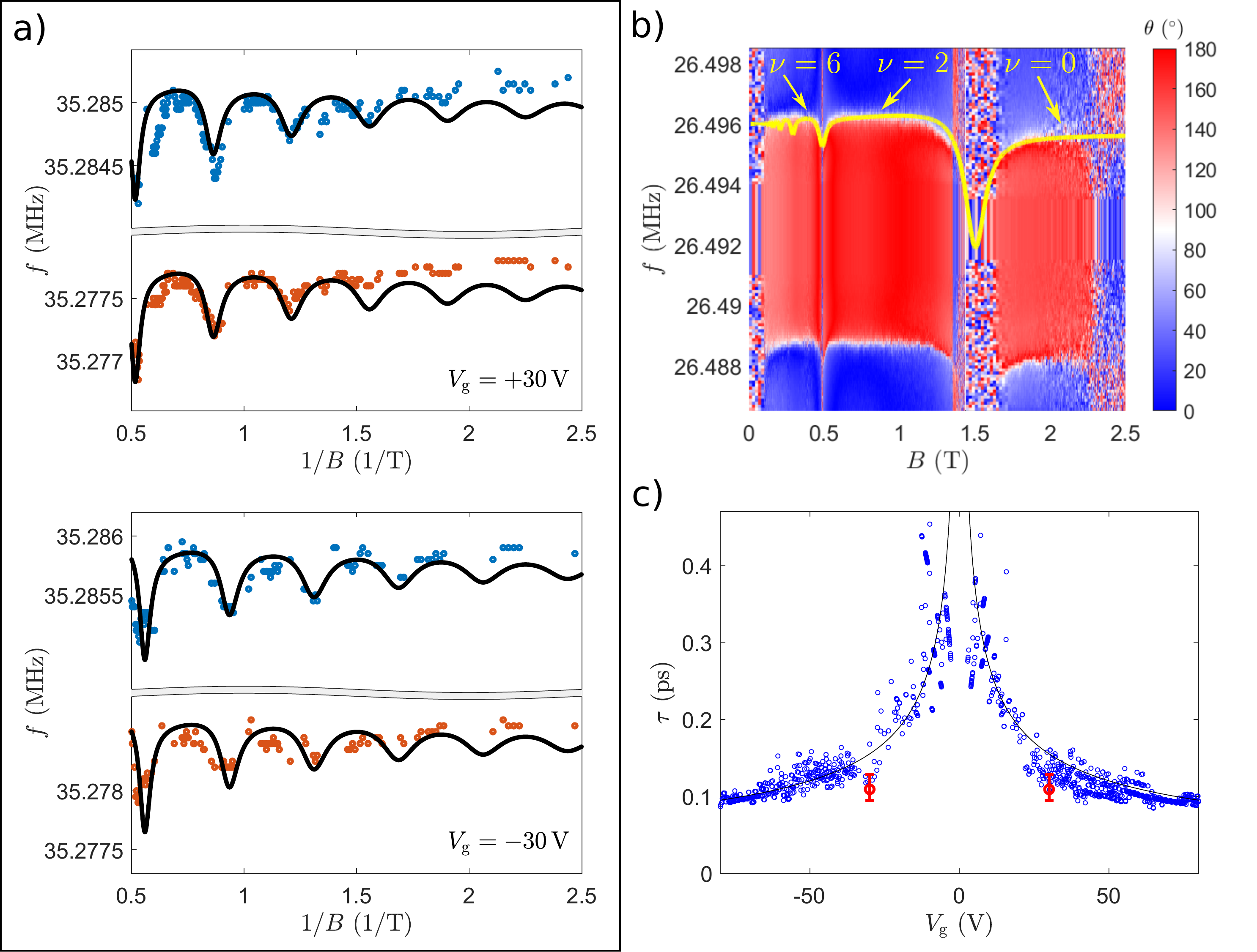}
\caption{{\bf Mechanical resonance frequency shift due to de Haas -- van Alphen effect.} 
  a) Upper (blue dots) and lower (red dots) edges of the $35\,$MHz 
  resonance in the B2 device at gate voltages $V_\mathrm{g} = \pm 30\,$V as a function 
  of $1/B$. The edge points correspond to the frequencies where the phase of the mechanically induced
  mixing current flips by 180 degrees (see Fig. \ref{fig:sample}). 
  The solid lines denote the theoretical fits with scattering times
  of $\tau_q \approx 0.11\,$ps at $V_\mathrm{g} = \pm 30\,$V. 
  b) Mixing current phase $(\theta)$ of the 26.5 MHz resonance in the
  device B1.5 presented as a function of $B$. The yellow line depicts
  the theoretical estimate with $\tau_q \approx 0.19\,$ps scattering time.
  c) Quantum scattering time $\tau_q$ extracted from dHvA
  measurement in Fig. \ref{fig:mainFreqShift}a (red markers), and the
  equivalent time $\tau_S$ from SdH oscillations in 
  Fig. \ref{fig:sample}a (blue markers). The red error bars show 
  a $15\%$ deviation from the chosen Landau level widths $\gamma$
  in Fig. \ref{fig:mainFreqShift}a that still reproduce a good 
  agreement between the theory and the experiment. 
  The solid black line denotes $\tau \propto 1/\sqrt{V_\mathrm{g}}$ trend.}
	\label{fig:mainFreqShift}
\end{figure}

In our measurements, we observed frequency shifts $\Delta f$, consistent with 
Eqs.~(\ref{eq:2},\ref{eq:6}), in the Au
resonator across the incompressible ranges of QH states.
Fig.~\ref{fig:mainFreqShift}a displays the magnetic field dependence of frequency 
points at which the measured phase of the mixing 
current flips $180^\circ$. The separation of these phase flips marks the linewidth of the resonance; 
the data were obtained in device B2 at $V_{\mathrm{g}}=\pm 30$V.  
The overlaid traces are calculated according to the theoretical
model for the dHvA effect given by Eq.~\eqref{eq:6}. The data indicate equivalent dHvA behavior for electrons and holes, which was also verified at other gate voltage values.
The extracted quantum scattering time
$\tau_q=\frac{\hbar}{\gamma}$ reduces as $V_\mathrm{g}$ increases, which is 
corroborated by a similar behavior in our other devices.
We resolve $\Delta f$ down to $\sim 25$ Hz, but for some magnetic field ranges, 
e.g. around $\nu = 2$ in Fig. \ref{fig:mainFreqShift}b, 
the magnitude of the frequency shift is not 
observable due to the low conductivity at the incompressible QH states. 
Another low-conductivity regime in Fig. \ref{fig:mainFreqShift}b is seen above $2.3\,$T, related to the 
state $\nu = 0$. In addition to the integer QH states, 
several fractional QH states are observed in these devices \citep{FQHE_corbino2021}, 
bearing witness to the sensitivity of our detection scheme.

It is worth noting that the theory predictions shown in Fig. \ref{fig:mainFreqShift}
are explicitly derived using the energy spectrum of massless Dirac 
electrons $\epsilon_n = \mathrm{sign}(n)\hbar\omega_\mathrm{D} \sqrt{\left| n \right|}$ 
implying a Berry phase $\gamma=\pm \pi$ \cite{Katsnelson2012}. 
Even though gold resonances are heavily utilized, we are probing 
the magnetization properties of the graphene part of the structure: 
Choosing the spectrum and Berry phase of 2D electron gas would 
result in a different spacing of the frequency dips.

In Fig.~\ref{fig:mainFreqShift}b, we present the phase of the mixing current measured in
the device B1.5 at the 26.5 MHz resonance at $V_\mathrm{g} = 7$ V as a function of perpendicular magnetic 
field. The overlaid curve is calculated from Eqs.~(\ref{eq:2},\ref{eq:6})
in which two fitting parameters were employed: $\partial 
f_\mathrm{0,n}/\partial V =35$~kHz/V and the quantum scattering time
$\tau_q=0.19$~ps. Equally good agreement is obtained for the lower phase flip as the width of 
the middle region (line width of the resonance) is unchanged across the measured magnetic field 
range. The fit-value of the scattering time is close to $\tau_S = $ 0.3 ps
extracted from Shubnikov--de Haas oscillations at $V_\mathrm{g} = 10$ V in a
similar device previously \cite{Kumar2018}, and also close to the data in Fig.~\ref{fig:mainFreqShift}c below. The general structure of the 
frequency modulation behavior $f(B)$ can be well understood by Eq.~\eqref{eq:2}, according to which 
the frequency dips are produced by the $-\frac{\partial f_\mathrm{0,n}}{\partial V}
\left[C_\mathrm{g} V_\mathrm{g} \Delta\frac{1}{C_\mathrm{q}} \right]$ 
term and the shift in the frequency level across the transition at $B=1.5\,$  from $\nu=2$ to $\nu=0$ 
arises from the ensuing $\Delta \mu \simeq 30$ meV (see Fig. S3). The last term in 
Eq.~\eqref{eq:2}, which would create positive peaks in the frequency shifts, is negligible with the 
parameters corresponding to our experiment. 

In Fig.~\ref{fig:mainFreqShift}c, the scattering times $\tau_q$
obtained from the dHvA fits are plotted as a function of $V_\mathrm{g}$ along
with the values $\tau_S$ obtained from the Shubnikov-de Haas
oscillations, present in the Landau fan plot in
Fig.~\ref{fig:sample}a. At small charge density, our value for
$\tau_S$ matches with the scattering time obtained in
Ref.~\cite{BOLOTIN2008} for ultraclean suspended graphene. The correspondence between the experimentally determined values of $\tau_q$ and $\tau_S$ corroborates the interpretation that our measured frequency shifts in $f(B,V_\mathrm{g})$, indeed arise from the dHvA effect in graphene.

\section{Outlook}

Compared to the other mechanical resonance measurements with graphene
samples in magnetic fields~\cite{Chen2016,Singh2012}, our approach is
different as we probe the graphene via a Au beam resonator. In a way
our work is similar to the cantilever experiments by Harris and
coworkers~\cite{Bleszynski-Jayich2009}: however, by using a ``second
cantilever'', the inner contact, we can facilitate operation on any
conducting 2D material and obtain extraordinary sample quality via
current annealing. This achievement seems out of reach for regular
cantilever devices combined with the present state-of-the-art
nanofabrication possibilities for 2D material. From this perspective, our setup opens up new possibilities in relation to the investigation of the magnetic properties of transition metal dichalcogenides (see e.g. \cite{Wang2017,Pisoni2018,Manzeli2019}), with particular reference to the role of local symmetry breaking in the appearance of magnetic moments and hidden interactions in centrosymmetric crystals \cite{Zhang.2014,Yuan2019,Du2021}. 

In conclusion, we have developed a versatile system of coupled resonators in
which a Au resonator can be employed for sensing of forces
originating in atomically thin suspended samples, made of graphene in the QH regime in our case. Owing to the free suspension of our graphene
membrane, movement of the Au sensing element can be detected via
displacement of the graphene, which facilitates force sensitivity
sufficient to observe magnetization oscillations due to the de
Haas--van Alphen effect in integer QH states, and even in the
fractional QH regime. The experimental approach developed in
this work opens up the possibility to investigate de Haas--van Alphen effect in other 2D materials, in particular transition metal dichalcogenide crystals with hidden magnetic properties.

\begin{acknowledgments}
  We thank V. Falko, M. Kumar, and S. Paraoanu for useful
  discussions. This work was supported by the Academy of Finland
  projects 314448 (BOLOSE) and 336813 (CoE, Quantum Technology
  Finland) as well as by ERC (grant no. 670743). This research project
  utilized the Aalto University OtaNano/LTL infrastructure which is
  part of European Microkelvin Platform. A.L.\ is grateful to
  Osk. Huttunen foundation for scholarship. J.M. thanks the support of the
  V{\"a}is{\"a}l{\"a} Foundation of the Finnish Academy of Science and
  Letters.
\end{acknowledgments}

\section{Methods}
\subsection{Sample fabrication and measurement setting}

Our Au resonators with suspended graphene Corbino disks were
fabricated using a method adapted from Ref.~\cite{Tombros2011} and
described in depth in Ref.~\cite{Kumar2018}. 
The fabrication was based on selective manipulation of the two
different resists with electron beam: PMMA for defining metal
contacts, and lift-off-resist (LOR) for support of suspended
structures. The Cr/Au contacts were deposited using ultra-high vacuum
metal evaporation in two steps with an LOR layer spun to separate the
bottom contact defining the outer rim of the Corbino device, while the
top contact supported by the separation layer bridged to the middle of
the Corbino ring. The thickness of the Au electrodes amounted
to $h_\mathrm{L}=70$ nm (the lower electrode) and $h_\mathrm{U}=120$ nm (the upper
electrode).

The whole structure was supported by $t_{\mathrm{V}} = 500$ nm
of LOR standing on a standard $t_{\mathrm{Si}}=300$ nm thick SiO$_2$ 
covered Si$++$ substrate that served as a back gate. Thus, the vacuum gap
(thickness $t_{\mathrm{V}}$) and SiO$_2$ as insulating layers yield 
$d_{\mathrm{eff}}=t_{\mathrm{V}}+t_{\mathrm{Si}}/\epsilon_{\mathrm{r}} = 580$ nm 
for the effective gap of the gate capacitor. This corresponds to the geometric gate capacitance value
$C_\mathrm{g} \approx 1.15 \times 10^{-5}$ F/m$^2$ obtained from the measured Landau fan diagram. Furthermore, the mechanical
resonant frequencies at $B=0$ exhibit the expected capacitive
softening behavior. Due to the built-in strain and a rather
large distance between the graphene and the gate electrodes, our
graphene membranes easily sustain voltages up to $V_\mathrm{g} = 100$ V. 

The fabricated sample chips were glued into sample boxes with
microwave striplines that in turn could be wire bonded to the bonding
pads of the Corbino devices. The sample boxes were connected to the
measurement lines through bias tees that allowed the DC conductance
measurements and low frequency readout of the mechanical resonance, as
well as the high frequency RF input used to transduce the mechanical
motion.

The results presented
here were obtained on two samples, B1.5 and B2, in which the Au beams
were connected graphene Corbino disks with outer (inner) diameter of
3.8 (1.5) and 4.5 (1.8) $\mu$m, respectively. The length of the main,
lower Au resonator, connected to the outer rim of graphene, was
approximately 8 $\mu$m, with a cross section of 70 nm $\times$ 5
$\mu$m.  The best results were obtained using resonance frequencies
around $25 - 35$ MHz, which corresponds to the third harmonic our
8-$\mu$m-long Au beam.  Quality factors of these resonances amounted
to $\sim 4000$.

The fabricated devices were characterized using standard conductance
measurement techniques and resonance measurements at 10 mK. The
devices were mounted slightly off center of the 9 T magnet on a
Bluefors LD400 dilution refrigerator. At the sample location,
$dB/dz=60$ T/m and $d^2B/dz^2= 1100$ T/m$^2$ with a maximum field of
6.8 T. This second derivative has such a tiny effect on the Au
mechanical frequencies so that it can be neglected in our force
analysis.

Prior to the actual measurements, however, current annealing
\cite{Moser2007} was performed by applying a bias voltage
$V_\mathrm{b} \approx 2$ V across the Corbino ring, consequently evaporating
residues from fabrication off from the graphene flake. The device
quality was assessed by measuring the Landau fan diagram, such diagram
is presented in Fig. \ref{fig:sample}a for the investigated sample
B2. Note the fractional QH state $\nu=1/3$ is visible from
$B\approx 3$ T upwards along with the usual set of integer quantum
Hall states highlighting the good quality of the measured samples. At
higher fields more fractional states appeared, see
Refs. \citep{Laitinen2018_halffilling,Kumar2018}.

\subsection{Identification of quantum Hall states}

Detection of the QH states in the graphene Corbino was performed both
using low-frequency AC conductance and the mechanical response of the
combined gold-graphene modes. The sensitive Au resonance
detection of QH states via graphene's mechanical response is
facilitated by the variation of the derivative $dG/dV_\mathrm{g}$ that
specifies the magnitude of the mixing current $I_\mathrm{mix}$ in
graphene (see Eq.~\eqref{eq:Imix}). Consequently, $I_\mathrm{mix}$ pinpoints
regions with $dG/dV_\mathrm{g}=0$, across which the mixing current
changes its sign. In the experiment, the sign change of
$I_\mathrm{mix}$ is seen as a flip of the phase by $\pi$ in the
down-mixed signal. Fig. \ref{fig:sample}c displays the measured
conductance $G(V_\mathrm{g})$ and the phase of the mixing current. An
exact match between $dG/dV_\mathrm{g}=0$ locations (traditional
location of the QH state) and the phase flips is observed.

\subsection{Mechanical resonances}

Mechanical resonances were detected using the FM mixing technique
\cite{Gouttenoire2010}. In this technique an FM-modulated signal
$V^\mathrm{FM}(t) = V^\mathrm{AC} \cos (2\pi f_\mathrm{c} t +(f_\Delta /f_\mathrm{L})\sin (2\pi f_\mathrm{L} t))$
was fed to the center electrode of the Corbino device through a bias
tee. Here $V^\mathrm{AC}$ and $f_\mathrm{c}$ are the carrier amplitude and frequency,
respectively. The sinusoidal low-frequency modulation signal at $f_\mathrm{L}$ (typically $\sim 600$
Hz) was supplied by the SR830 lockin amplifier, while the
frequency deviation $f_{\Delta}$ (typically 1-4 kHz) was produced by
the frequency generator (Rohde \& Schwarz SMY01, or Keysight N9310A)
producing the FM-modulated signal.

The FM-modulated signal, applied across source-drain electrodes of the
graphene membrane, got downmixed by the intrinsic nonlinearity of the
graphene device, and the low frequency component at the frequency
$f_\mathrm{L}$ reflects the mechanical motion amplitude $z$ of the graphene
flake. This proportionality can be expressed as
\begin{equation}
I_\mathrm{mix} \propto \frac{C'_\mathrm{tot}}{C_\mathrm{tot}} \frac{\partial G}{\partial V_\mathrm{g}} \left| \frac{\partial \mathrm{Re}(z)}{\partial f} \right|,
\label{eq:Imix}
\end{equation}
where $C_\mathrm{tot} = (1/C_\mathrm{g}+1/C_\mathrm{q})^{-1}$ is the total capacitance, and $C'_\mathrm{tot}=\frac{dC_\mathrm{tot}}{dz}$. 
$C_\mathrm{g}$ and $C_\mathrm{q}$ are the gate
capacitance and the quantum capacitance per unit area, respectively.
Phase shifts may occur between the drive and the response due
interference phenomena in the flexural waves traveling along the
Corbino disk, driven from the outer edge.
In the case of phase shifts, the mechanical response function
$\frac{\partial \mathrm{Re}(z)}{\partial f}$ will obtain a corresponding
reference phase, which results in a combination of dispersive and
absorptive parts of the mechanical response.

The observed combined gold-graphene modes below $\sim 40$ MHz
involve either the lower or upper Au electrode beam, the motion of
which is followed by graphene at master-slave principle owing to the
time-dependent boundary conditions imposed by Au on graphene. 
The $35.3\,$MHz resonance of the B2 device depicted
in Fig. \ref{fig:sample}d, for example, is detected with the
measurement configuration shown in Fig. \ref{fig:schematic} but is not
observable when the source and drain sides are reversed implying that
this mechanical mode is dominated by the movement of the lower gold
beam with the graphene sheet following.
Moreover, it is this exclusive reliance on the mechanical Au
resonances in actuation and detection, regardless of the properties of
graphene, that allows us to generalize our investigation method to
other 2D materials.

In our frequency sweeps of the sample B1.5, we observed 12 mechanical
modes below 27 MHz. Using COMSOL simulations, candidate mode shapes
for these modes could be identified.
We utilized simulated gate voltage dependencies for each mode to
determine the mode shape corresponding to the 26.5 MHz resonance with
which the dHvA effect was observed. A mode shape, where the most
significant role is played by the cantilever, displays a weak
frequency increase with respect to the gate voltage in the simulations
corresponding to the observed trend in the measurements.

We emphasize that our detection scheme for magnetization effects in a 2D 
material relies on finding well-defined resonances of the gold structure 
and is, therefore, suitable for a very wide variety 2D systems. Specifically,
the 2D-material portion of the structure is not required to have good resonator properties. For example, in our case, the resonances shown here have quality factors of the order of 1000.

In addition to graphene and Au modes, surface waves around 20 MHz
were excited in the LOR-layer by the microwave drive
\cite{Laitinen2019_lamb}. Even though these modes could be excited at
very small power, they were not useful for detection purposes owing to
the small $Q =100 - 200$.

The sensitivity of our experiments is set by the frequency resolution
of the resonance peak position, approximately 25 Hz. This frequency
resolution corresponds to $\sim 10^4$ Bohr magnetons, which is six orders
of magnitude better than in the torque magnetometer work of
Ref. \cite{Wiegers1997}. 
Compared with the cantilever
work of Ref.~\cite{Harris1999}, our sensitivity is two orders of
magnitude better. After optimization of the device parameters and
improving the frequency resolution, similar sensitivity as in the work
of Bleszynski-Jayich et al. can be obtained
\cite{Bleszynski-Jayich2009}.

\subsection{Fitting the theory predictions}

The theoretical predictions of the frequency shift due to the dHvA 
effect given in Eq.~\eqref{eq:2} are fitted to the experimental data as 
shown in Figs. \ref{fig:mainFreqShift}a and \ref{fig:mainFreqShift}b. To obtain the spring constant 
of the modes, we assume that the effective mass of the modes is 
determined by the gold resonator, and graphene's contribution is 
practically negligible. We approximate that, for the B2 device $\sim 80\,$nm 
thick lower gold beam, the 2D density is $\rho \approx 1.5\times10^{-3}\,$kg/m$^2$, 
and $\rho \approx 2.1\times10^{-3}\,$kg/m$^2$ for the $\sim 110\,$nm thick
cantilever of B1.5.

For our setup, the last term of the frequency shift in Eq. 
\eqref{eq:2} is 
negligible and, therefore, the size of the frequency shift is scaled by 
the factor $\partial f_\mathrm{0,n}/\partial V$. We fit $\partial 
f_\mathrm{0,n}/\partial V$ to the experimental data to obtain proper 
magnitude of $\Delta f$ together with the Landau level width $\gamma$ 
that affects the magnitude of $\Delta f$ as well as the width of the 
frequency dips. For the device B2 (B1.5) we have $16
\,$kHz/V ($35\,$kHz/V). The larger value for the B1.5 
device mode is expected due to it being a mode of the cantilever whose 
other end is attached to the graphene. This boundary condition makes 
the B1.5 mode more sensitive to tensioning effects than the resonance 
of the lower gold plate of the B2 device.

The scattering time $\tau_S=\hbar \sqrt{\pi n} \mu_q/e v_\mathrm{F}$ shown 
in Fig. \ref{fig:mainFreqShift}c was calculated from the
quantum mobility $\mu_q$, which in turn was obtained by extracting the
minimum field of Shubnikov--de Haas oscillations $B_0(V_\mathrm{g})$ and using
the relation $\mu_q B_0 = 1$. Additionally, the error bars of $\tau_\mathrm{q}$ in 
the same figure show a $15\%$ deviation from the values of $\gamma = \hbar/(2 \tau_\mathrm{q})$ 
used to fit the theory curves in Fig. \ref{fig:mainFreqShift}a. Values of 
$\gamma$ within this tolerance reproduce a good agreement of $\Delta f$ 
between the theory and the experiment.

\subsubsection{Supplementary Information}
\label{sec:supp}

Details of the theoretical results.

\bibliographystyle{apsrevsimpl}
\bibliography{references}

\end{document}


\title{De Haas--van Alphen effect in graphene --
  Supplementary material}

\author{Juuso Manninen}
\thanks{These two authors contributed equally}
\affiliation{Low Temperature Laboratory, Department of Applied
  Physics, Aalto University School of Science, P.O. Box 15100,
  00076 Aalto, Finland}
\affiliation{QTF Centre of Excellence, Department of Applied Physics, Aalto University, PO Box 15100, FI-00076 AALTO, Finland}
\author{Antti Laitinen}
\thanks{These two authors contributed equally}
\affiliation{Low Temperature Laboratory, Department of Applied
  Physics, Aalto University School of Science, P.O. Box 15100,
  00076 Aalto, Finland}
\affiliation{Department of Physics, Harvard University, Cambridge, MA 02138, USA}
\author{Francesco Massel}
\email[]{francesco.massel@usn.no}
\affiliation{Department of Physics, Nanoscience Center, University of Jyv\"{a}skyl\"{a}, FIN 40014, Finland}
\affiliation{Department of Science and Industry Systems, University of
  South-Eastern Norway, PO Box 235, Kongsberg, Norway}

\author{Pertti Hakonen}
\affiliation{Low Temperature Laboratory, Department of Applied
  Physics, Aalto University School of Science, P.O. Box 15100,
  00076 Aalto, Finland}
\affiliation{QTF Centre of Excellence, Department of Applied Physics, Aalto University, PO Box 15100, FI-00076 AALTO, Finland}

\maketitle

\section{Thermodynamic analysis  and forces acting on a movable capacitor (finite $C_\mathrm{q}$)}
Let us consider the energy per unit area of a capacitor with one of the plates
constituted by a finite-DOS material (in our case graphene), in the
presence of an external magnetic field $B$
\begin{equation}
\label{eq:1}
    U \left(n, B  \right) = \frac{1}{2} \frac{e^2 n^2}{C_\mathrm{g}} + \Xi \left( n,B \right). 
\end{equation}
The first term in Eq.~\eqref{eq:1} is the energy
associated with the electrical field building up between the plates of
the capacitor, whereas the second corresponds to the energy related to
the finite density of states (DOS) of the system. In this description, the
charge $n$ on the capacitor plates and the magnetic field $B$ are the
control parameters.

\begin{figure}
\centering
\includegraphics[width=0.5\textwidth]{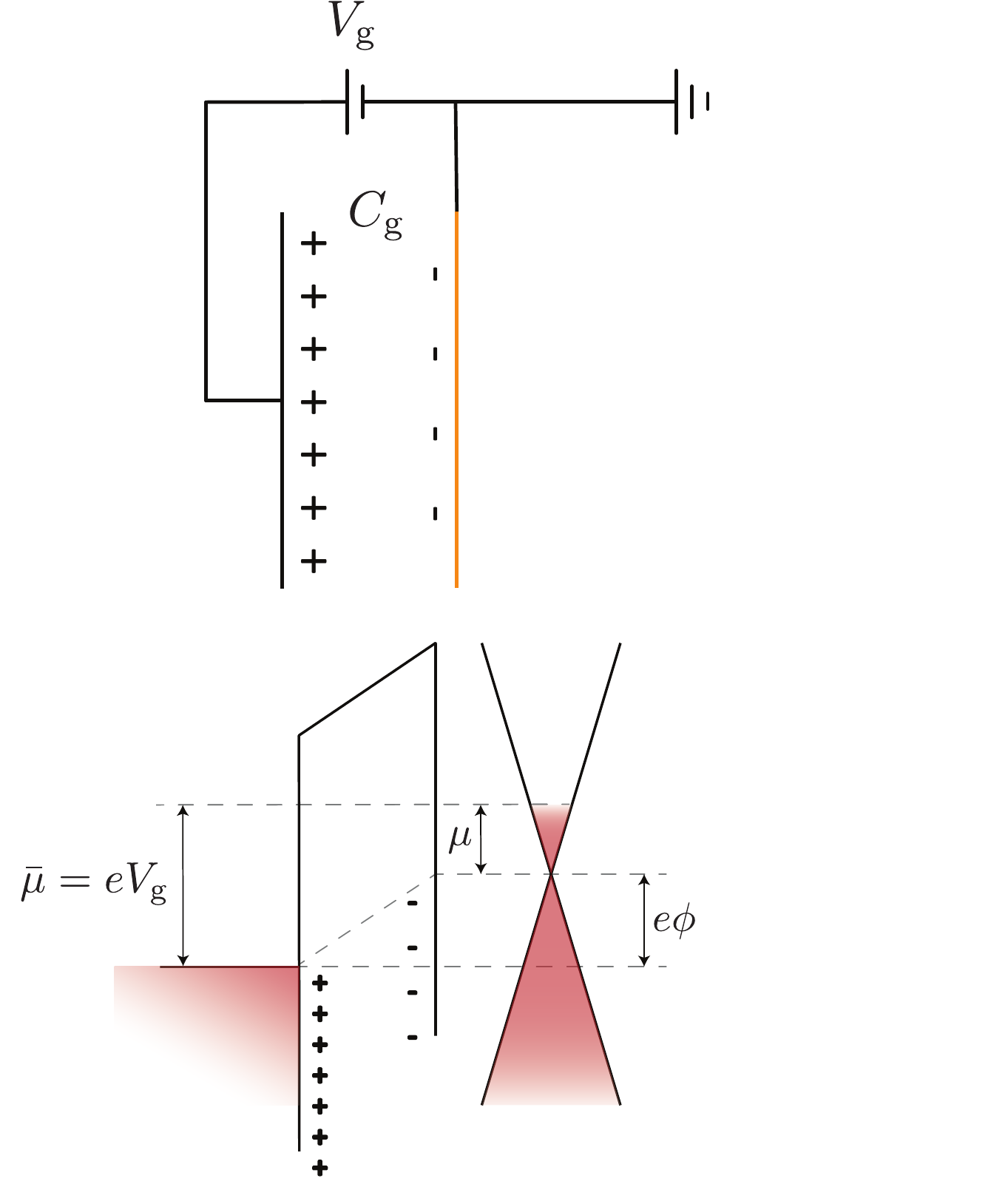}
\caption{Top. Lumped-elements description of a capacitor with one
  plate constituted by a finite-DOS material (yellow) to which a voltage
  $V_\mathrm{g}$ is applied. Bottom. The voltage drop leads to the
  buildup of an electric field between the capacitor plates,
  corresponding to an electrostatic potential drop $\phi$. Owing to
  the finite DOS, part of
  the energy provided by the voltage source goes into promoting
  electrons to higher-energy single-particle states (increase of the
  chemical potential $\mu$).  
}
\label{fig:Vdrop}
\end{figure}
Considering a standard thermodynamic relation~\cite{Callen.2006}, we
can define the electrochemical potential as
\begin{equation}
\label{eq:2}
\bar{\mu} = \left( \frac{\partial U}{\partial  n} \right)_{z,B} =
 \frac{e^2 n}{ C_\mathrm{g}} + \mu \left( n \right) 
\end{equation}
implying
$n = \frac{C_\mathrm{g}}{e^2} \left( \bar{\mu} - \mu \right)$. This relation, written as
$$
\frac{\bar{\mu}}{e} = \frac{e n}{C_\mathrm{g}} + \frac{\mu}{e},
$$ 
allows us to show that, for $C_\mathrm{g} \to \infty$, the tunability of the electrochemical potential $\bar{\mu}$ is directly translated into the tunability of the chemical potential of the membrane $\mu$.
Here $\bar{\mu}$ is the electrochemical potential consisting of the
electrostatic $e \phi = \frac{e^2 n}{C_\mathrm{g}}$ and chemical
$\mu = \frac{\partial \Xi}{\partial n}$ potentials~\cite{Yu.2013}.

We can now consider the thermodynamic potential $\Omega \left( \bar{\mu} \right)$
where $\bar{\mu}$ (not $\mu$ nor $\phi$) is the independent control
parameter, along with $B$. From equation~\eqref{eq:1}, we obtain
\begin{equation}
\label{eq:3}
    \Omega \left( \bar{\mu},B\right) = U \left( n, B \right) - \bar{\mu} n = \frac{1}{2} \frac{C_\mathrm{g}}{e^2} \left( \bar{\mu} - \mu \right)^2 + \Phi \left( \bar{\mu}, z \right) - \bar{\mu} n ,
\end{equation}
where we have defined
$\Phi \left( \bar{\mu}, B \right) \doteq \Xi \left( n \left( \bar{\mu}, B \right) \right)$.
Using equation~\eqref{eq:3}, we can calculate the force for
constant $\bar{\mu}$, i.e. at constant external applied voltage
\begin{align}
  V_\mathrm{g} \doteq \frac{\bar{\mu}}{e} = \frac{e n}{C_\mathrm{g}} + \frac{\mu}{e}.
  \label{eq:4}
\end{align}
To this end, let us obtain a preliminary result
\begin{equation}
\label{eq:5}
\begin{split}
    \frac{\partial n}{\partial  z} &= {C_\mathrm{g}'}{e} \left( V_\mathrm{g} - \frac{\mu}{e} \right) - \frac{C_\mathrm{g}}{e^2} \pdv{\mu}{n} \pdv{n}{z} = \frac{C_\mathrm{g}'}{e} \left( V_\mathrm{g} - \frac{\mu}{e} \right) - \frac{C_\mathrm{g}}{C_\mathrm{q}} \frac{\partial  n}{\partial z}  \\
    \implies \frac{\partial n}{\partial z} &= \left( 1 + \frac{C_\mathrm{g}}{C_\mathrm{q}} \right)^{-1} \frac{C_\mathrm{g}'}{ e} \left(V_\mathrm{g} - \frac{\mu}{e} \right),
\end{split}
\end{equation}
where $C_\mathrm{g}=\partial C_\mathrm{g} / \partial z$. Equation~\eqref{eq:5} implies, in addition, that
\begin{align}
  \label{eq:6}
  \frac{\partial \mu}{\partial z} &=  \frac{C_\mathrm{g}'}{\left(C_\mathrm{q} + C_\mathrm{g} \right)}
                                      \left(e V_\mathrm{g} - \mu \right)
\end{align}
and that the electrostatic force per unit area can be expressed as
\begin{equation}
\label{eq:7}
\begin{split}
    F &= - \frac{\partial \Omega}{\partial  z} = - \frac{1}{2} C_\mathrm{g}' \left( V_\mathrm{g} - \frac{\mu}{e} \right)^2 + \frac{C_\mathrm{g}}{e} \left(V_\mathrm{g} - \frac{\mu}{e} \right) \pdv{\mu}{n} \pdv{n}{z} - \mu \pdv{n}{z} + e V_\mathrm{g} \frac{\partial n}{\partial z} \\
    &= - \frac{1}{2} C_\mathrm{g}' \left(V_\mathrm{g} - \frac{\mu}{e} \right)^2 + \frac{C_\mathrm{g}}{C_\mathrm{q}} e \left( V_g - \frac{\mu}{e} \right) \frac{\partial n}{\partial  z} + e \left(V_\mathrm{g} - \frac{\mu}{e} \right) \frac{\partial n}{\partial z} \\
    &= - \frac{1}{2} C_\mathrm{g}' \left( V_\mathrm{g} - \frac{\mu}{e} \right)^2 \left( 1 - 2 \frac{1 + \frac{C_\mathrm{g}}{C_\mathrm{q}}}{1 + \frac{C_\mathrm{g}}{C_\mathrm{q}}} \right) \\
    &= \frac{1}{2} C_\mathrm{g}' \left( V_\mathrm{g} - \frac{\mu}{e} \right)^2, 
\end{split}
\end{equation}
which is consistent with the expression given in the literature about
carbon nanotubes~\cite{Steele2009,Lassagne2009}.

\begin{figure}[htb]
\centering
  \includegraphics[width=\textwidth]{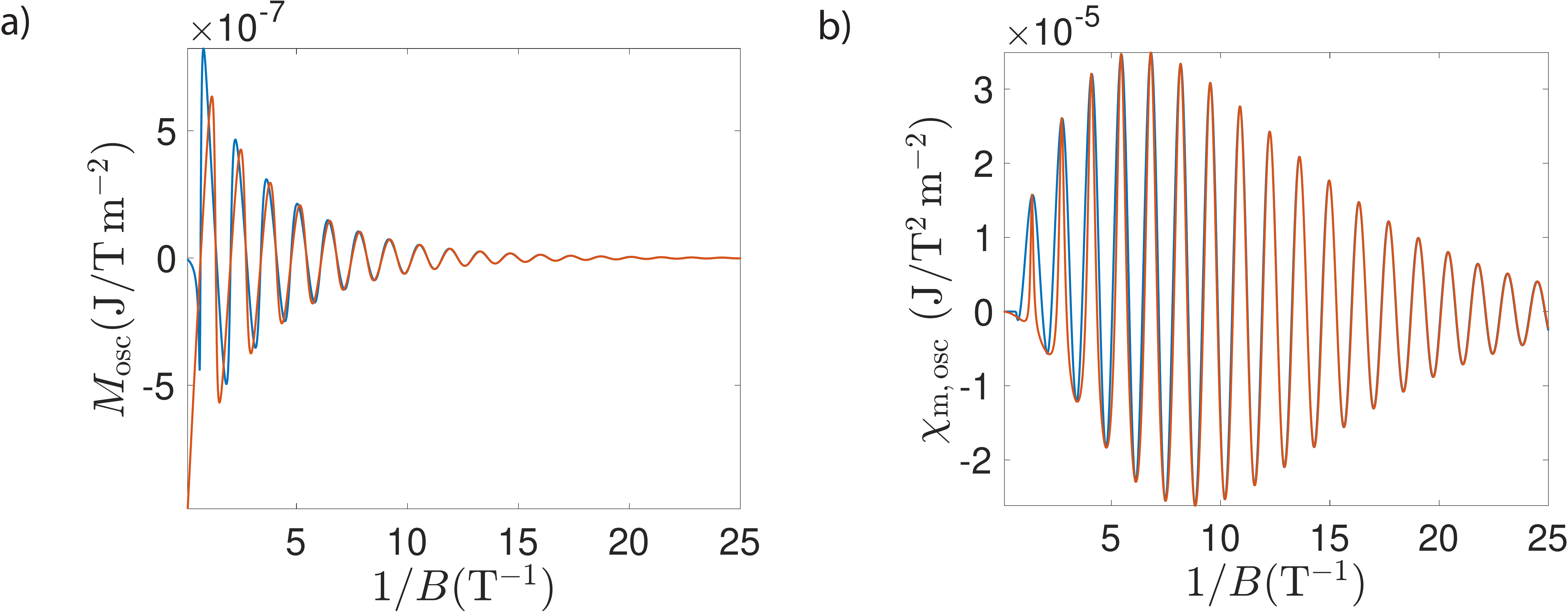}
	\caption{a) Magnetization oscillations obtained from equation~\ref{eq:3} using
          $v_\mathrm{F}=10^6$ m/s and $\tau_q= 0.19$ ps 
          at fixed $\mu = 31$ meV (red) and $\mu (B)$ for $C_\mathrm{g} \approx 1.15 \times 10^{-5}$ F/m$^2$   (blue),
          where the Lorentzian broadening with the same disorder has
          been taken into account.  b) The oscillating part $\chi_\mathrm{m,osc}$ of magnetic susceptibility
          $\chi_m (\mu,B)$ for the same parameters as in panel a).}	
	\label{fig:M_theory}
 \end{figure}

As discussed in the main text, the distinguishing factor between the full thermodynamic 
potential $\Omega \left( \bar{\mu},B \right)$ and that of the graphene disk 
$\Omega_\mathrm{disk} \left( \mu,B \right)$ is the control parameter $\bar{\mu}$ 
versus $\mu$. In Fig.~\ref{fig:M_theory}, we have compared the behavior of the oscillatory 
components of magnetization $M_\mathrm{osc}$ and magnetic susceptibility $\chi_\mathrm{m,osc}$, 
see the derivations below, as a function of $B^{-1}$ with $\bar{\mu}$ and $\mu$ as a control
parameter.

\section{Frequency Shift}
\label{sec:frequency-shift}

Given the expression for the force of Eq.~\eqref{eq:7} (Eq.~(1) of
the main text), we evaluate here the mechanical frequency shift for a
mechanical resonator whose electrical properties can be described by
Eq.~\eqref{eq:1}.

For sake of simplicity, we focus here on the analysis of a membrane
(vanishing flexural rigidity) in the presence of uniform tension. The
only difference with the general case (finite flexural rigidity and
nonuniform tension) is the specific value of the mechanical frequency
in absence of external applied voltage $\omega_\mathrm{0,n}$, which is
not central to our argument.

Firstly, we notice that the distance between the
plates of the capacitor --and, consequently, the capacitance
$C_\mathrm{g}$-- is modified by the application of the external
electrostatic potential $V_\mathrm{g}$ (see Fig.~\ref{fig:Vdrop}). The
equilibrium position of the movable capacitance
$\xi_\mathrm{e}=\xi_\mathrm{e}\left(V_\mathrm{g},B\right)$ can be
obtained from the solution of the general elastic equation for the
structure (see e.g. \cite{Atalaya.2008} for the case of a
graphene-only membrane).

The shifts of the mechanical resonant frequency $\omega_\mathrm{0,n}$
for $V_\mathrm{g}\neq 0$ can be determined considering the membrane
position fluctuations around $\xi_\mathrm{e}$. For a membrane, these
fluctuations obey the following equation
\begin{align}
  \rho \ddot{\zeta} - T \Delta \zeta =  \delta f(\mathrm{r},t) ,
  \label{eq:8}
\end{align}
where we have defined
$ \delta f = \left[f(\xi_\mathrm{e}+\zeta) -f(\xi_\mathrm{e})\right]$ 
and $T=T(\xi_\mathrm{e})$.
The functions $\zeta(\mathrm{r},t)$ and $\delta f(\mathrm{r},t)$ in Eq.~\eqref{eq:8} 
can be expanded as
\begin{align}
  \zeta(\mathrm{r},t) = \sum_\mathbf{n} \zeta_\mathrm{n} \exp\left[- i  \omega_\mathrm{n} t \right] A_\mathrm{n}(\mathrm{r}) ,  \nonumber\\
  \delta F(\mathrm{r},t) = \sum_\mathbf{n} \delta F_\mathrm{n} \exp\left[- i  \omega_\mathrm{n} t \right] A_\mathrm{n}(\mathrm{r}) , \nonumber
\end{align}
where $A(\mathrm{r})$ are the normal modes associated to the boundary
value problem considered. Furthermore, approximating 
$$   \delta F_\mathrm{n}   \simeq  \left. \frac{\partial F }{\partial \zeta}\right|_{\xi_\mathrm{e}} \zeta_\mathrm{n}, $$
we obtain
\begin{align}
  \rho \omega_\mathrm{n}^2 \zeta_\mathrm{n}
      - T \lambda_\mathrm{n}^2 \zeta_\mathrm{n} = \left. \frac{\partial F}{\partial \zeta}\right|_{\xi_\mathrm{e}} \zeta_\mathrm{n},
  \label{eq:10}
\end{align}
 leading to the following expression for the frequency of a given mode
$\mathrm{n}$
\begin{align}
  \omega_\mathrm{n}=\sqrt{\omega_{0,\mathrm{n}}^2-\frac{1}{\rho} \left. \frac{\partial F}{\partial \zeta}\right|_{\xi_\mathrm{e}}} 
  \label{eq:11}
\end{align}
with $\omega_{0,\mathrm{n}}=\sqrt{T \lambda^2_\mathrm{n}/\rho},$ where
$\lambda_\mathrm{n}$ depends on the geometry considered. For the case
of a disk of radius $R$, we have that 
$\lambda_\mathrm{n} \to  \lambda_{i,j} = \alpha_{i,j} / R$, where
$ \alpha_{i,j} $ are the roots of the Bessel function $J_{i,j}(r)$.
For the case of metallic leads (infinite DOS), we have that $\partial F/\partial \zeta)=1/2 C_\mathrm{g}'' V_\mathrm{g}^2$ allowing us to obtain from Eq.~\eqref{eq:11} the usual capacitive softening voltage dependence of the mechanical resonant frequency. 

In the case of finite DOS, from Eq.~\eqref{eq:7}, we can write
\begin{align}
  \left.\frac{\partial F}{\partial \zeta}\right|_{\xi_\mathrm{e}}  =
  \left.\frac{1}{2} C_\mathrm{g}'' \left(V_\mathrm{g}-\frac{\mu}{e}\right)^2 - C_\mathrm{g}' \left(V_\mathrm{g}-\frac{\mu}{e}\right)
  \frac{1}{e}\frac{\partial \mu}{\partial \zeta}\right|_{\xi_\mathrm{e}} ,
  \label{eq:12}
\end{align}
which with the help of Eq.~\eqref{eq:6}, gives
\begin{align}
  \left. \frac{\partial F}{\partial \zeta}\right|_{\xi_\mathrm{e}} =
 \left. \left\{ \frac{1}{2} C_\mathrm{g}'' - \frac{{C_\mathrm{g}'}^2}{C_\mathrm{q}+C_\mathrm{g}} \right\}\left(V_\mathrm{g}-\frac{\mu}{e}\right)^2\right|_{\xi_\mathrm{e}}.
  \label{eq:14}
\end{align}
Denoting by $\Delta \omega_\mathrm{n}$ the difference between the resonant frequency at finite $B$ with respect to the frequency at $B=0$, we can write
\begin{align}
  \Delta \omega_\mathrm{n} \simeq \Delta \omega_{0,\mathrm{n}}-
  \frac{1}{2 \rho \omega_{0,\mathrm{n}}}
  \left(\left.\frac{\partial F}{\partial \zeta}\right|_{B}-
        \left.\frac{\partial F}{\partial \zeta}\right|_{B=0}\right) ,
  \label{eq:15}
\end{align}
where $\Delta \omega_{0,\mathrm{n}}=\omega_\mathrm{0,\mathrm{n}}(B)-\omega_\mathrm{0,\mathrm{n}}(B=0)$. The first term in Eq.~\eqref{eq:15} represents the change of the resonant frequency with the external magnetic field associated with tensioning effects, whereas the second term is related to the magnetic field change of the capacitive softening term.

Noting that, for $B=0$, the resonant frequency $\omega_{0,\mathrm{n}}$ depends on the externally
applied voltage only, we can approximate
\begin{align}
  \Delta \omega_{0,\mathrm{n}}=& \left.\frac{\partial \omega_{0,\mathrm{n}}}{\partial V}
                                 \frac{\partial V}{\partial F}\right|_{B=0} \Delta F
                             \simeq  - \left.\frac{\partial \omega_{0,\mathrm{n}}}{\partial V}\right|_{B=0}
                                \frac{\Delta \mu}{e}.
                          \label{eq:16}
\end{align}
Integrating equation~\eqref{eq:6}, in the limit
$V_\mathrm{g}\gg \mu/e$ and $C_\mathrm{q} \gg C_\mathrm{g}$, we have
\begin{align}
   \mu(B,\xi) = \mu(B) + \frac{e\, C_\mathrm{g}(\xi)V_\mathrm{g}}{C_\mathrm{q}(B)},
  \label{eq:17}
\end{align}
where the first term corresponds to the change in chemical potential
as a function of $B$ in the absence of capacitive detection
($C_\mathrm{g}=0$).

Equations~(\ref{eq:16},~\ref{eq:17}) lead to
\begin{align}
  \Delta \omega_{0,\mathrm{n}}=&-\frac{\partial \omega_{0,\mathrm{n}}}{\partial V}
                             \left[
                                 \Delta \mu + C_\mathrm{g} V_\mathrm{g} \Delta\left(\frac{1}{C_\mathrm{q}}\right)
                             \right].
  \label{eq:18}
\end{align}
Analogously, we have that
\begin{align}
  \left.\frac{\partial F}{\partial \zeta}\right|_{B}-\left.\frac{\partial F}{\partial \zeta}\right|_{B=0} \simeq
  -{C''}_\mathrm{g} V_\mathrm{g} \left[\frac{\Delta \mu}{e}+ C_\mathrm{g} V_\mathrm{g} \Delta\left(\frac{1}{C_\mathrm{q}}\right) \right] - \left({C'}_\mathrm{g} V_\mathrm{g}\right)^2
  \Delta\left(\frac{1}{C_\mathrm{q}}\right), 
  \label{eq:19}
\end{align}
leading to 
\begin{align}
   \Delta \omega_{\mathrm{n}}=\left[-\frac{\partial \omega_{0,\mathrm{n}}}{\partial V}+\frac{C'' V_g}{2 \rho \omega_{0,\mathrm{n}}}\right]\left[\frac{\Delta \mu}{e}+ C_\mathrm{g} V_\mathrm{g} \Delta\frac{1}{C_\mathrm{q}} \right]+\frac{\left(C'_\mathrm{g} V_\mathrm{g}\right)^2}{2\rho \omega_{0,\mathrm{n}}}\Delta\left(\frac{1}{C_\mathrm{q}}\right),
  \label{eq:20}
\end{align}
which with $\kappa_\mathrm{n}=\rho \omega_\mathrm{n}^2$ and
$\omega_\mathrm{n}=2 \pi f_\mathrm{n}$ leads to Eq.~(2) of the
main text. We note also that, in the limit $C_\mathrm{g}''=0$, 
Eq.~\eqref{eq:20} corresponds to the expression given in
\cite{Chen.2015} for the mechanical frequency shift.

\section{Quantum capacitance}

\begin{figure}
\centering
\includegraphics[width=0.5\textwidth]{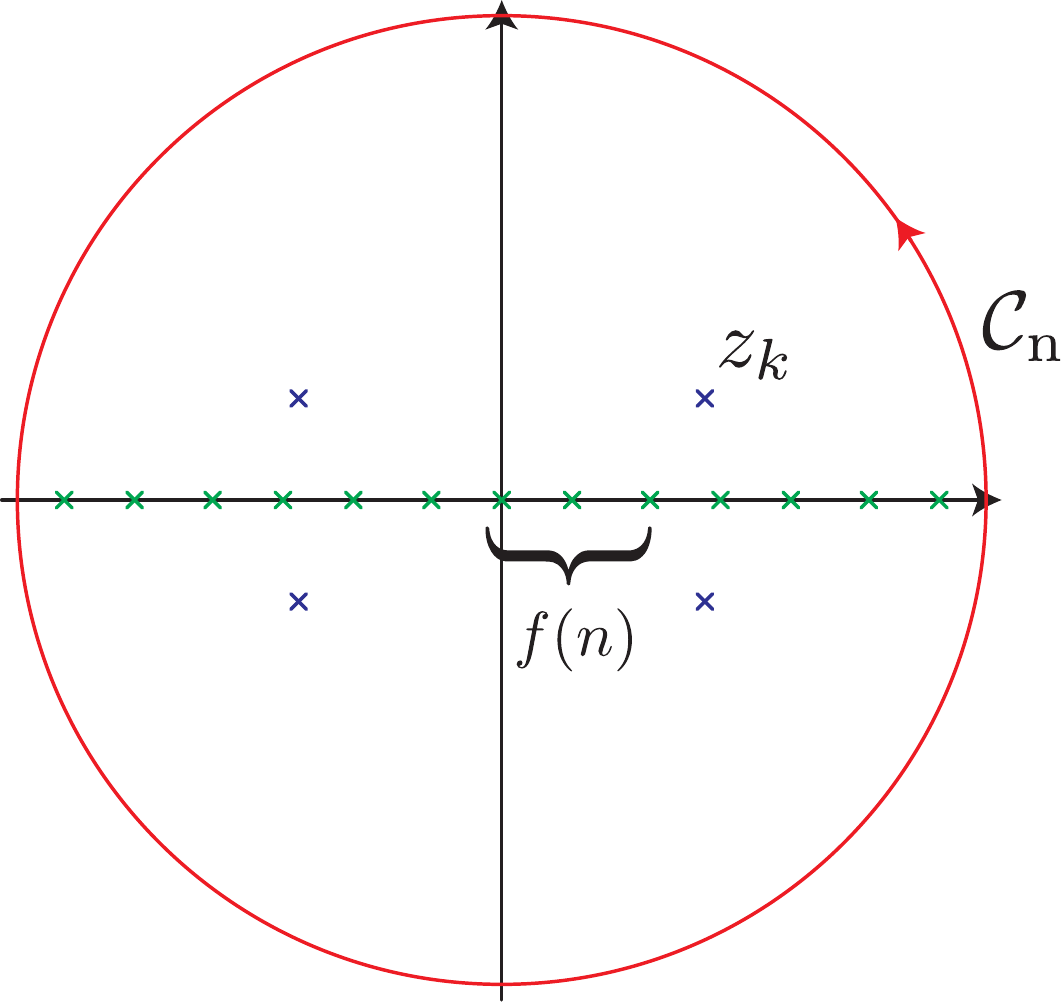}
\caption{Complex-plane representation of the relation in
  Eq.~\eqref{eq:24}. The red line represents the contour along which
  the integral on the lhs is performed. We have denoted the poles of
  $ 2 \pi z \cot(\pi z)$ (whose residue is $f(n)$) in blue, whereas
  the poles of  $f\left(z^2\right)$ are marked in red.}
\label{fig:Contour}
\end{figure}

We derive here the expression for the quantum capacitance
$C_\mathrm{q}$.  To this end, we express the density of states as a
sum of Lorentzians centered at
$\epsilon_\mathrm{n}=\mathrm{sign}(n)\hbar \omega_\mathrm{D} \sqrt{|n|}$
($\omega_\mathrm{D}=v_\mathrm{f}\sqrt{2 e B / \hbar}$)
\begin{align}
  \label{eq:21}
  D(\epsilon)= \frac{N e B}{2 \pi \hbar} \sum^\infty_{n=-\infty} \frac{\gamma}{2 \pi} \frac{1}{\left(\epsilon-\epsilon_n\right)^2+ \frac{\gamma^2}{4}}
\end{align}
with the spin degeneracy factor $N=2$.
Our analysis is based on the possibility of turning an infinite sum
into an integral over the complex plane.  
If $f(w)$ is a meromorphic function, the following condition is fulfilled
\begin{align}
  \sum f(n) =\oint_{\mathcal{C}_\mathrm{n}}  \pi \cot(\pi w) f(w) d w -\sum_k \mathrm{Res}\left[\pi \cot(\pi w) f(w);w_k\right].
  \label{eq:22}
\end{align}
Since, in our case, we have
\begin{align}
  f(n)=\frac{N e B}{2 \pi \hbar} \frac{\gamma}{2 \pi}\frac{1}{\left(\omega-\mathrm{sign}(n)\omega_\mathrm{D}\sqrt{|n|}\right)^2+\frac{\gamma^2}{4}}
  \label{eq:23}
\end{align}
we operate a change of variables $w \to z^2$, which allows us to
rewrite Eq.~\eqref{eq:22} as
\begin{align}
  \label{eq:24}
   \sum f(n) =\oint_{\mathcal{C}_\mathrm{n}} 2 \pi z \cot(\pi z) f(z^2) d z -\sum_k \mathrm{Res}\left[2 \pi z \cot(\pi z) f\left(z^2\right);z^2_k\right] ,
\end{align}
see Fig.~\ref{fig:Contour} for the complex-plane representation of
contours and poles leading to Eq.~\eqref{eq:24}.

The integral on the lhs of Eq.~\eqref{eq:24} represents a (diverging)
constant contribution, which can be renormalized by introducing an
explicit frequency cutoff, analogously to what is done in
Ref.~\cite{Sharapov2004}. Conversely, the second term, the sum over
the poles of $2 \pi z \cot(\pi z) f(z)$, can be evaluated explicitly,
leading to the following expression for the density of states 
\begin{align}
  D(\epsilon)=\Lambda_D + \frac{N}{\pi v_F^2 \hbar^2} \frac{\epsilon \sinh\left[\frac{2 \pi \gamma \epsilon}{(\hbar \omega_\mathrm{D})^2}\right]-\frac{\gamma}{2}\sin\left[\frac{2 \pi \left(\epsilon^2-\frac{\gamma^2}{4}\right)}{(\hbar \omega_\mathrm{D})^2}\right]}{\cosh\left[\frac{2 \pi \gamma \epsilon}{(\hbar \omega_\mathrm{D})^2}\right]-\cos\left[\frac{2 \pi \left(\epsilon^2-\frac{\gamma^2}{4}\right)}{(\hbar\omega_\mathrm{D})^2}\right]} ,
  \label{eq:25}
\end{align}
where $\Lambda_\mathrm{D}$ is the contribution coming from the
integral on the contour $\mathcal{C}_\mathrm{n}$.

In the $\omega\gg \gamma,\, \omega_\mathrm{D}$ limit the expression
for $D(\omega)$ given in Eq.~\eqref{eq:25}, coincides with the one
given in Ref.~\cite{Sharapov2004}, in the same limit. From
Eq.~\eqref{eq:25} we can easily derive the expression for
$C_\mathrm{q}= e^2 D(\mu)$.

\section{Quantum capacitance and de Haas -- van Alphen effect}

With an analogous calculation to the one leading to the expression for
$C_\mathrm{q},$ it is possible to derive an expression for the charge
density $n$
\begin{align}
   n=n_\mathrm{0}+n_{\mathrm{osc}}
  \label{eq:26}
\end{align}
with 
\begin{subequations}
\begin{align}
  \label{eq:27}
  n_\mathrm{0}&=\frac{N}{2\pi v_F^2 \hbar^2}\left(\mu^2-\frac{\gamma^2}{4}\right)\\
  n_{\mathrm{osc}}
                          &=  \frac{N e B}{\pi^2 \hbar} \arctan \left\{ \frac{\sin \left[ \frac{2 \pi}{(\hbar \omega_\mathrm{D})^2} \left( \mu^2 - \frac{\gamma^2}{4} \right) \right]}{\exp \left[ \frac{2 \pi \gamma \mu}{(\hbar \omega_\mathrm{D})^2} \right] - \cos \left[ \frac{2 \pi}{(\hbar \omega_\mathrm{D})^2} \left( \mu^2 - \frac{\gamma^2}{4} \right) \right]} \right\}
\end{align}
\end{subequations}
and to the oscillatory component of the magnetization
$M_{\mathrm{osc}}=-\frac{\partial \Omega_\mathrm{osc}}{\partial B}$
\begin{align}
  \label{eq:28}
  M_\mathrm{osc} &= 
   -\frac{N e }{2 \pi^2 \hbar} \mu \arctan \left\{ \frac{\sin \left[ \frac{2 \pi}{(\hbar \omega_\mathrm{D})^2} \left( \mu^2 - \frac{\gamma^2}{4} \right) \right]}{\exp \left[ \frac{2 \pi \gamma \mu}{(\hbar \omega_\mathrm{D})^2} \right] - \cos \left[ \frac{2 \pi}{(\hbar \omega_\mathrm{D})^2} \left( \mu^2 - \frac{\gamma^2}{4} \right) \right]} \right\} .
\end{align}
From Eqs.~(\ref{eq:27},~\ref{eq:28}) is straightforward to
establish a relation between the oscillating component of the quantum capacitance
$C_\mathrm{q,osc}=e^2 \partial n_\mathrm{osc}/\partial \mu$
and the oscillations of the magnetic susceptibility
$\chi_\mathrm{m,osc}=\frac{\partial M_\mathrm{osc}}{\partial B}$. From
Eqs.~(\ref{eq:27}-\ref{eq:28}), we have that
\begin{subequations}
\begin{align}
  \label{eq:29}
  C_\mathrm{q,osc}&=-\frac{N e^2}{\pi v_\mathrm{f}^2 \hbar^2}
  \frac{
     \mu  \left\{
       e^{- 2 \pi \gamma \mu /\hbar \omega_\mathrm{D}}-\cos\left[\frac{2 \pi}{(\hbar \omega_\mathrm{D})^2} \left( \mu^2 - \frac{\gamma^2}{4} \right)\right]
  \right\}
   +\gamma/2
  \sin\left[
  \frac{2 \pi}{(\hbar \omega_\mathrm{D})^2}\left( \mu^2 - \frac{\gamma^2}{4} \right)
  \right]
       }
       {
       \cosh \left[ \frac{2 \pi \gamma \mu } {\left(\hbar \omega_\mathrm{D}\right)^2}\right]
       -\cos\left[\frac{2 \pi}{(\hbar \omega_\mathrm{D})^2}\left( \mu^2 - \frac{\gamma^2}{4} \right)\right]
       } , \\
  \label{eq:30}
   \chi_\mathrm{m,osc} &= -\frac{N e  \mu} {4 \pi \hbar B}
    \frac{}{}
  \frac{\left( \mu^2 - \frac{\gamma^2}{4} \right)
   \left\{
       e^{- 2 \pi \gamma \mu /\hbar \omega_\mathrm{D}}-\cos\left[\frac{2 \pi}{(\hbar \omega_\mathrm{D})^2} \left( \mu^2 - \frac{\gamma^2}{4} \right)\right]
  \right\}
  +\gamma \mu
  \sin\left[
  \frac{2 \pi}{(\hbar \omega_\mathrm{D})^2}\left( \mu^2 - \frac{\gamma^2}{4} \right)
  \right]
  }
  {
  (\hbar \omega_\mathrm{D})^2 \left\{\cosh \left[ \frac{2 \pi \gamma \mu } {\left(\hbar \omega_\mathrm{D}\right)^2}\right]
 -\cos\left[\frac{2 \pi}{(\hbar \omega_\mathrm{D})^2}\left( \mu^2 - \frac{\gamma^2}{4} \right)\right]\right\}
  } ,
\end{align}
\end{subequations}
which leads to the relation 
\begin{align}
  \label{eq:31}
   C_\mathrm{q,osc}= \frac{\chi_\mathrm{m,osc}} {\Gamma(\mu,B)} 
\end{align}
with
\begin{align}
  \Gamma(\mu,B) =\frac{\mu^2-\gamma^2/4}{4 B^2 e^2}
  +\frac{\gamma}{4 B^2 e^2}\frac{\mu^2+\gamma^2/4}
        {\gamma+2 \mu  \sin\left[\frac{2 \pi}{(\hbar \omega_\mathrm{D})^2}\left( \mu^2 - \frac{\gamma^2}{4} \right) \right]^{-1} \left\{\cos\left[\frac{2 \pi}{(\hbar \omega_\mathrm{D})^2}\left( \mu^2 - \frac{\gamma^2}{4} \right) \right]- \exp\left[\frac{2 \pi \gamma \mu}{  (\hbar \omega_\mathrm{D})^2}\right]\right\}}.
  \label{eq:32}
\end{align}
Here for $\mu \gg \gamma$, we obtain
\begin{align}
  \Gamma(\mu,B) =\left(\frac{\mu}{2 e B}\right)^2 .
  \label{eq:33}
\end{align}

\begin{figure}[htb]
\centering
\includegraphics[width=0.95\textwidth]{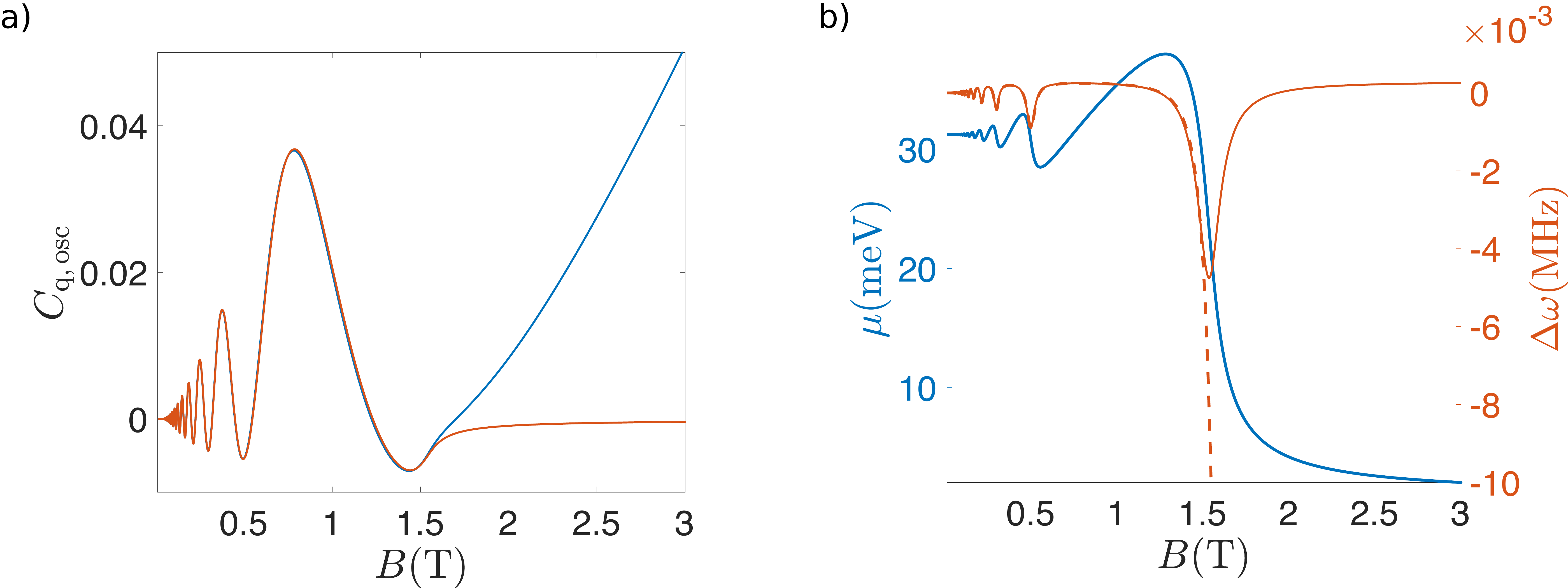}
\caption{a) Value of $C_\mathrm{q,\,osc}$ calculated through the
  relation
  $C_\mathrm{q,osc}= \frac{\chi_\mathrm{m,osc}} {\Gamma(\mu,B)}$, for
  the value of $\Gamma(\mu,B)$ given in equation \eqref{eq:32} (blue line),
  and for the zeroth-order expansion in $\gamma$  (equation \eqref{eq:33}, red line).  b) Corresponding plot of the frequency shift. Exact value of $\Gamma(\mu,B)$
  (continuous line), zeroth-order expansion in $\gamma$ (dotted
  line). All parameters are the ones used for the data fit of device B1.5. 
}
\label{fig:gamma}
\end{figure}

Neglecting the small oscillations of the chemical potential, it is
therefore clear that the dips in the dependence of the mechanical
resonant frequency $\omega_\mathrm{n}$ as a function of $B$, which
were previously been interpreted in terms of quantum capacitance
oscillations, can equivalently be interpreted in terms of oscillations
of the magnetic susceptibility $\chi_\mathrm{m}$ (de Haas -- van Alphen effect).

\newpage

\bibliographystyle{apsrevsimpl} 
\bibliography{references_supp}